\documentclass[12pt]{article}
\usepackage{amsmath}
\renewcommand{\thefootnote}{\fnsymbol{footnote}}
\newfont{\bit}{cmbxti10 scaled 1728}
\begin{document}
\newpage
\thispagestyle{empty}
\begin{center}
{\LARGE {ADM and Bondi four-momenta\\
 for the \\
ultrarelativistic Schwarzschild black hole\\ 
}}

\vspace{2cm}
{\large
 Peter C. AICHELBURG 
 \footnote[1]{ e-mail: pcaich@doppler.thp.univie.ac.at}
}\\
{\em 
 Institut f\"ur Theoretische Physik, Universit\"at Wien\\
 Boltzmanngasse 5, A - 1090 Wien, AUSTRIA 
 }\\[.5cm]
{\em and}\\[.5cm]
{\large Herbert BALASIN
\footnote[2]{e-mail: hbalasin@tph.tuwien.ac.at}
}\\ 
{\em
 Institut f\"ur Theoretische Physik, Technische Universit\"at Wien\\
 Wiedner Hauptstra{\ss}e 8--10, A - 1040 Wien, AUSTRIA 
}\\[.8cm]
\end{center}

\begin{abstract}
We argue that it is possible to assign Bondi as well as ADM
four-momentum to the ultrarelativistic limit of the
Schwarzschild black hole  in agreement to what is expected on physical 
grounds: The Bondi-momentum is lightlike and equal to the ADM-momentum up to 
the retarded time when particle and radiation escape to infinity and 
drops to zero thereafter, leaving flat space behind. 
\end{abstract}
\vfill

\rightline{UWThPh -- 1998 -- 12}
\rightline{TUW 98 -- 08}
\rightline{March 1998}

\newpage
\small\normalsize
\renewcommand{\thefootnote}{\arabic{footnote}}
\setcounter{footnote}{0}
\newpage 
\pagebreak
\pagenumbering{arabic}
\pagestyle{plain}

\section*{\Large\bit Introduction}
Black holes, specifically the Schwarzschild geometry, are prototypes 
of isolated systems. This intuitive concept is mathematically captured in 
the notion of asymptotic flatness \cite{Walker}, which states that 
``at large distances'' the geometry approaches flat space. Precisely this 
notion of asymptotic Minkowski space allows to define asymptotic symmetries, 
specifically asymptotic translations. Its corresponding charges represent the 
total energy-momentum at spatial infinity and the energy-momentum left in the 
system at a given moment of retarded time at null infinity. For the 
Schwarzschild geometry, ADM and Bondi momenta agree, since there is 
no radiation due to the static nature of the spacetime. \\
In the present article we consider the ultrarelativistic limit
of the Schwarz\-schild black hole \cite{AS}, which is an impulsive, 
plane-fronted wave with parallel rays \cite{JEK}. 
It represents the gravitational field as it 
appears to an asymptotic observer moving with ultrarelativistic speed 
(in fact at the speed of light) relative to black hole. 
In this limit the gravitational field becomes completely concentrated on 
a null hyperplane, spacetime being flat above and below the pulse.
A closer look at the wave profile reveals that it decays logarithmically 
along spacelike directions in the null plane thus leaving little hope 
for asymptotic flatness.
On the other hand from its construction as limiting boost of 
Schwarzschild it seems physically reasonable to assign the 
corresponding ultrarelativistic (null) limit of the 
ADM momentum to the AS-geometry. At this point it is important to note that 
both geometries belong to the Kerr-Schild class, which is characterized by a 
geometric decomposition of the metric into a flat part and the tensor square 
of a geodetic null vector-field. Moreover, the limiting process can
be considered as a one-parameter family of geometries within this class
\cite{BaNa3}. We will make use of the flat part
of the decomposition to define asymptotic translations. Application 
of this procedure to the Schwarzschild geometry produces
the correct, well-known result of a constant timelike Bondi and
ADM momentum. Encouraged by this we extend the procedure
to the limit geometry: A careful distributional treatment produces
a lightlike ADM momentum. Similar calculations, taking the limit in 
lightlike directions, give a Bondi momentum equal to ADM up
to the instant of retarded time where the pulse reaches infinity.
For later retarded times the Bondi momentum is zero, showing that
the energy was completely radiated away. The physical picture just described
is a direct result of the aforementioned calculation.
It is also possible to use (generalized) conformal techniques
\cite{Penrose2,Ashtekar}, which endow asymptotically flat spacetimes
with certain boundaries and allow to interpret the energy four-momenta 
as integrals over two-dimensional cross-sections of these boundaries.
However, we will refrain from the presentation of this construction 
in the present essay,
reserving the full calculations for a subsequent publication \cite{AiBa6}.

\section*{\bit Energy integrals for the ultrarelativistic \\
Schwarzschild black hole}

To give an illustration of our approach let us start from the 
Schwarzschild geometry
\begin{align*}
&g_{ab} = \eta_{ab} + f \>k_a k_b\qquad k^ak^b\eta_{ab} = 0\quad
(k\nabla)k^a = (k\partial)k^a =0\\
&f=\frac{2m}{r}\quad k^a = \partial_t^a + \partial_r^a
\end{align*}
which has been written in a form that explicitly exhibits it as
a member of the Kerr-Schild class.
The coordinates $t,r,\theta,\phi$ denote standard spherical
polar coordinates in Minkowski space.
Arbitrary translations with respect to the flat part
of the decomposition $\alpha^a$, $\partial_b\alpha^a=0$, 
are asymptotically Killing, as can be seen from
$$
\nabla^{(a} \alpha^{b)} = C^{(b}{}_c{}^{a)} \alpha^c =
\frac{1}{2}((\alpha\partial)(fk^a k^b) - f(k\partial)f(\alpha k)\> k^a k^b), 
$$
which behaves like $1/r^2$. The Bondi four-momentum is defined by
the Komar expression \cite{Walker}, which evaluates the curl of $\alpha^a$ 
on a two-sphere moving along a surface of constant retarded
time to null infinity 
\begin{align*}
&P^{Bondi}_a \> \alpha^a = \lim_{v\to \infty}
\frac{1}{2}\int_{S^2_{vu}} \nabla^{[a} \alpha^{b]}\epsilon_{abcd}=
-8\pi m\alpha^t,
\end{align*}
whereas the ADM four-momentum is given by an Ashtekar-Hansen type 
expression \cite{Chrus}
\begin{align*}
&P^{ADM}_a \> \alpha^a = \lim_{r\to \infty}
\frac{1}{2}\int_{S^2_{tr}}\!\! R_{abmn}\alpha^a x^b\epsilon^{mn}{}_{cd}
= -8\pi m\alpha^t,
\end{align*}
where by a slight abuse of notation the Lie-Algebra of translation has 
been identified with its representation as vector-fields.
Here and in the following $u,v$ will denote the retarded and advanced 
coordinates associated with $t,r$.
The result reflects the staticity of the gravitational field, i.e. the 
absence of gravitational waves carrying away mass-energy.
Let us now turn to the ultrarelativistic version of the Schwarzschild
field. This geometry belongs to a subfamily of the Kerr-Schild class, 
the so-called plane-fronted gravitational waves with parallel rays 
(pp-waves), characterized by the existence of a covariantly constant
vector field $p^a$
\begin{align*}
&g_{ab} = \eta_{ab} + f\> p_a p_b\qquad p^a p^b\eta_{ab}=0\quad 
\nabla_a p^b =\partial_a p^b =0\quad (p\partial)f=0\\
&f=\mu \delta(t-z)\log\rho\quad p^a = \partial_t^a + \partial_z^a 
\end{align*}
$\rho,z$ are related to $r,\theta$ like cylindrical to spherical
polar coordinates. Due to the impulsive nature of the wave profile, i.e. 
$f\sim \delta(t-z)$  the spacetime is flat almost everywhere. 
We will therefore once again use the translations relative to the flat 
part of the decomposition to calculate the Bondi four-momentum 
as its corresponding curl,
\begin{align*}
&\nabla^{[a} \alpha^{b]} = (p\alpha) \partial^{[a} f p^{b]}, \qquad
\frac{1}{2}\nabla^{[a} \alpha^{b]}\epsilon_{abcd} =
(p\alpha)\mu\delta(t-z) p_{[c} d\phi_{d]},\\
\\
&P^{Bondi}_a \> \alpha^a = \lim_{v\to \infty}
\frac{1}{2}\int_{S^2_{vu}} \nabla^{[a} \alpha^{b]}\epsilon_{abcd}\\
&\hspace*{1.6cm}=(p\alpha) \lim_{v\to\infty}\int \mu 
\delta(\frac{u}{2}(1+\cos\theta) + \frac{v}{2}(1-\cos\theta))
\frac{v-u}{2} \sin\theta d\theta d\phi \\
&\hspace*{1.6cm}= 2\pi\mu\> \Theta(-u)\> p_a\alpha^a, \\
\end{align*}
where $\Theta$ denotes the Heaviside function. \newline
A closer look at the 
integrand of the ADM-expression 
\begin{align*}
&R_{abcd}\alpha^a x^b &&= \frac{1}{2}((x\partial)\partial_d f p_c 
-(x\partial)\partial_c f p_d)(p\alpha) -
((\alpha\partial)\partial_d f p_c (\alpha\partial)\partial_c f p_d)(px))\\
&&&=-\frac{1}{2}(p\alpha)(p_c \partial_d f - p_d \partial_d f),
\end{align*}
which made use of the negative homogeneity of $\partial_a f$ and 
the identity $(px)\delta'(px)= - \delta(px)$, reveals that it 
coincides with the Komar expression. Therefore
\begin{align*}
P^{ADM}_a \> \alpha^a &= \lim_{t\to \infty}
\frac{1}{2}\int_{S^2_{tr}} \nabla^{[a} \alpha^{b]}\epsilon_{abcd}\\
&=(p\alpha) \lim_{t\to\infty}\int \mu 
\delta(t-r\cos\theta)r\sin\theta d\theta d\phi \\
&=2\pi\mu\> p_a\alpha^a.
\end{align*}
It is precisely the negative homogeneity of the $\delta$-function, i.e. 
$\delta(\lambda x) = \lambda^{-1} \delta(x)\quad \lambda > 0$, which
guarantees the convergence of the above integrals.
This is also the reason why the slow logarithmic decay on the null
hyperplane does not spoil the asymptotic behavior. 

\section*{\bit Concluding remarks}

Since the gravitational field of the 
ultrarelativistic Schwarzschild geometry is 
concentrated on a null hyperplane it necessarily distributional in
nature. We were able to show that this 
property is responsible for the convergence of the  
integrals of the total energy and the energy left in the system
after the emission of gravitational radiation. Viewed in this
context the above geometry is a nice example for a system  that completely
radiates off all energy leaving flat space behind.
Due to its close connection with the Schwarzschild black hole the
limit geometry plays a distinguished role within the class of impulsive
pp-waves. Waves with a profile proportional to higher multipoles
are asymptotically flat but have zero energy-momentum \cite{BeigChrus}. 
A closer look reveals, however, that they necessarily violate the
(weak) energy condition. 
On the other hand for pp-waves that do not fall off,
e.g. plane waves, the energy-momentum is undefined.
Finally, we would like to comment on the theorem that the ADM and Bondi 
momenta cannot be null \cite{AshHor,BeigChrus}.
In this context it is important to remark that the spacetime, due to its 
distributional nature, does not obey the conditions prerequisite to this 
conclusion, i.e. that it represents singular initial data. 
Nevertheless, we do believe that the simple physical
interpretation of the result shows that this geometry represents
a sensible (idealized) physical system, which is asymptotically
flat in a distributional setting. 
Actually, the situation is reminiscent of the 
symmetry classification of pp-waves by Jordan Ehlers and Kundt, which
assigns the ultrarelativistic Schwarzschild geometry a two-parametric
symmetry group. However, precisely the distributional nature of 
the profile shows that the symmetry group is four-parametric as it is 
for Schwarzschild \cite{AiBa1}. \\
\vfill

\noindent
The authors thank R.~Beig and B.~Schmidt for helpful discussions,
and P.~Chrusciel for a remark concerning the ADM-expression.\\
This work was supported in part by the Fundacion Federico.

\end{document}